\documentclass{aa}
\usepackage{graphicx}
\usepackage{txfonts}  
\newcommand {\be}{\begin{equation}}
\newcommand {\ee}{\end{equation}}
\newcommand {\bea}{\begin{eqnarray}}
\newcommand {\eea}{\end{eqnarray}}

\begin{document}

\title{High efficiency of soft X-ray radiation reprocessing in supersoft X-ray
sources due to multiple scattering}
\titlerunning{High efficiency of soft X-ray radiation reprocessing}
\author{V. Suleimanov \inst{1,3} \and
F. Meyer \inst{2}
\and E. Meyer-Hofmeister \inst{2}}
\offprints{V. Suleimanov, \\ \email{vals@ksu.ru}}

\institute{%
Department of Astronomy, Kazan State University, Kremlevskaya 18, Kazan 8,
Russia, 420008
\and Max-Planck-Institut
f\"ur Astrophysik, Karl-Schwarzschild-Stra{\ss}e 1,
 D-85740 Garching, Germany
\and Kazan Branch of Isaac Newton Institute, Santiago, Chile}

\date{Received  / Accepted }

\abstract{
Detailed analysis of the lightcurve of CAL 87 clearly has shown
that the high optical luminosity comes from the accretion disc rim
and can only be explained by a severe thickening of the disc rim
near the location where the accretion stream impinges. This area
is irradiated by the X-rays where it faces the white dwarf.
Only if the reprocessing rate of X-rays to optical light is high
a luminosity as high as observed can be understood. But a recent detailed study
of the soft X-ray radiation reprocessing in supersoft X-ray sources
has shown that the efficiency is not high enough.
We here propose a solution for this problem. As already discussed in
the earlier lightcurve analysis the impact of the
accretion stream at the outer disc rim produces a ``spray'',
consisting of a large number of individual gas blobs imbedded in a
surrounding corona. For the high mass flow rate this constitutes an
optically thick vertically extended screen at the rim of the accretion disc.
We analyse the optical properties of this irradiated spray and find that
the multiple scattering between these gas blobs leads to an effective
reprocessing of soft X-rays to optical light as required by the observations.
}
\maketitle

\section {Introduction}
The recent progress in X-ray observations provides us with a wealth of
X-ray spectra for a continuously increasing number of objects. A special
class of objects are the supersoft X-ray sources, X-ray binaries with a
very soft spectrum (most photons have an energy below 0.5 keV) and a
very high luminosity ($10^{37}$ to $10^{38}$ erg/s).
These binaries are commonly
understood as a white dwarf in a close orbit with a low-mass star
where the mass flow towards the compact star forms an accretion disc.
The very high accretion rate ($10^{-7}$M$_\odot$/yr) causes steady state
burning on the white dwarf surface. The high accretion rate also is
responsible for a thickening of the accretion disc rim as found to a
lower amount also for the discs in low mass X-ray binaries (Milgrom
\cite{Milgrom}, Mason \& Cordova \cite{MC}, White et al. \cite{WNP}).
A detailed analysis of the lightcurve of CAL 87 ( Schandl,
Meyer-Hofmeister \& Meyer \cite{SMM}) had already
shown that the high optical luminosity can only be explained by a severe
thickening of the disc rim where the accretion stream impinges. This area
is irradiated by the X-rays where it faces the white dwarf.
A high reprocessing rate of X-rays to optical light is needed
to produce a luminosity as high as observed. A recent detailed study
of the soft X-ray radiation reprocessing in supersoft X-ray sources
(Suleimanov, Meyer \& Meyer-Hofmeister \cite{SlMM}) has shown that the
efficiency is not high enough.
We now here discuss a new model. We suggest that
due to the high mass flow rate a ``spray'' exists, a vertically extended
area around the impact of the accretion stream at the outer
disc rim. This area is filled with gas blobs.
In the presented paper we discuss the optical properties of gas blobs
and show how the light is reprocessed in such a spray.

In Sect. 2 we discuss
the qualitative picture of the  soft X-ray reprocessing during multiple
scattering between gas blobs.
In Sect. 3 the
situation simplified to the radiation
transfer in a flat cloud slab is considered.
In Sect. 4 we desribe the application  of the presented model to CAL 87
based on the lightcurve simulation of Schandl et al. (\cite{SMM}). A
discussion and conclusions follow in Sect. 5 and Sect. 6.

\section{Reprocessing of soft X-ray to optical radiation by multiple
scattering between gas clouds}

In previous work (Suleimanov, Meyer \& Meyer-Hofmeister \cite{SlMM}) we showed
that the reprocessing efficiency $\eta$ is very small (0.05 - 0.1) if an accretion
disc is irradiated by a soft X-ray flux ($E <$ 1-2 keV). Due to the high
opacity in this spectral band the soft X-rays can not
penetrate deeply into the disc atmosphere. As a result the soft X-ray
flux heats the upper atmospheric layers and is
reradiated in the far UV spectral band. The temperature of deeper layers,
where the optical disc radiation is formed, is increased only slightly. It is
obvious that the same qualitative picture holds for an optically
thick gas cloud with a temperature $T_c \sim 10^4$~K,
irradiated by the soft X-ray flux. It means that the reprocessing
efficiency $A$ due to a single  reradiation from the cloud irradiated by
the soft X-rays is small too.
But the efficiency $\eta$ can be  higher for a slab consisting of many separate
gas clouds, irradiated by the soft X-rays.

We denote by $A$ the fraction  of
incoming  soft X-ray/far UV flux that is reradiated in the
optical spectral band due to reradiation at a single cloud. By $\eta$ we
denote the fraction of incoming soft X-ray/far UV flux that
is reradiated in the optical spectral band due to reradiation from
an accretion disc or a cloud slab.  We use the term "soft X-ray" for the
spectral band with photon energy 0.1-1 keV and "far UV" for the spectral
band with photon energy 0.015 - 0.1 keV.

The opacity of a plasma with $T \approx 10^4$~K is high in the far UV
band. If an optically thick cloud is irradiated by a far UV
flux this flux can not penetrate deeply into the cloud. The far UV flux
heats the upper cloud layers and is reradiated in the same far UV spectral
band.  The temperature of deeper layers, where the cloud optical radiation
is formed, is increased only slightly. Therefore, the reprocessing
efficiency $A$ for the far UV flux  to the optical band is expected to be
small for reradiation at a single cloud.  Thus reprocessing efficiencies
in the soft X-ray and far UV band should be very similar, and thus
below we will consider a common soft X-ray/far UV band.

In the stationary case all flux impacting on a cloud  must be
reradiated.  As the electron scattering cross section in a cloud with $T
\approx 10^4$~K is small compared to the absorption cross section we
may assume that the incoming soft X-ray/far UV flux is fully
absorbed, heats the upper cloud layers, and then is reradiated. We refer
to this process as  reflection.  But the reradiated flux is not a black
body radiation with a single temperature.  The part reradiated in the far
UV band has higher radiation temperature then the part reradiated in
the optical band.

For a single reflection from a gas cloud, the main part, $(1-A)F_x$,
of the  absorbed soft X-ray flux $F_x$ is reradiated in the far UV
band, only a small part, $AF_x$, is reradiated in the
optical band.  As mentioned above $A$, the reprocessing efficiency for
single reflection, is around 0.05-0.1.  Therefore,
a reflection from a cloud can be considered as a kind of scattering.
But this is a not scattering of a photon in the sense of an
elementary physical process. It is only sutable description  of
the process of flux reflection from one cloud, which makes it simpler
to describe the radiation transfer through a slab of clouds. If reflected
from a slab consisting of many individual clouds radiation can have many
such "scatterings" between clouds before it exits from the slab again,
resulting in a reprocessing efficiency   that is higher than $A$. In the
next section we show that this efficiency $\eta$ can reach up to 0.5 (with
$A$=0.1) in the most favourable case.

\begin{figure}
\resizebox{88mm}{!}{\includegraphics{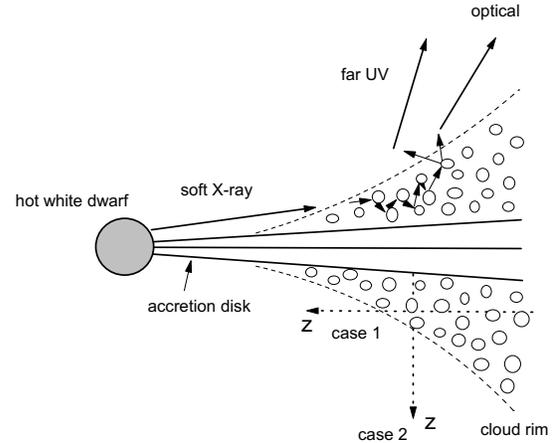}} \vspace{-20mm}
\caption[]{The qualitative picture of the cloud rim above an accretion disc in
supersoft X-ray sources.}
\label{ql_pic}
\end{figure}

The optical light curve   of the supersoft X-ray source CAL 87 was
modelled by Schandl, Meyer-Hofmeister \& Meyer (\cite{SMM}) with the
assumption of $\eta = 0.5$. They showed that the outer parts of
the accretion disc must be extended in z-direction.
This extended outer rim is necessary to intercept a significant part of
the hot white dwarf flux and to reprocess it into the optical band. In that
work it was also shown that this rim can not be homogeneous and the suggestion
was made that the rim consists of separate clouds, which can be a spray
formed at the impact of the gas stream from the secondary star
onto the disc. What we call ``clouds'' here are geometrically small
cloudlets or gas blobs.
The cloud radii should be much smaller than the disk radius in agreement
with their formation on the scale of a two-stream-instability in the shear
interaction. Thus, we suggest the following qualitative picture of the outer
parts of the accretion disk in supersoft sources (see Fig. \ref{ql_pic}).
Above the outer geometrically thin disk there is a slab of gas clouds
embedded into a hot ($T \approx 5 \cdot 10^5$ K) intercloud medium.
This geometrically thick slab intercepts a significant part from the
soft X-ray radiation of the central source. This radiation is scattered
many times between clouds inside the slab and finally reflected from the
slab. We discuss the physical parameters of the clouds and the intercloud
medium in Sect. 4.   In a single scattering a
small fraction (0.05 -0.1) of the soft X-ray/far UV flux is
reradiated in the optical band, but due to many scatterings before
exit the final reprocessing efficiency increases by several
times and can reach up to 0.5. This model can
explain the high value of the soft X-ray reprocessing observed in some
supersoft sources, such as CAL 83 and CAL 87.

\section{Radiation transfer in the gas cloud slab}

\begin{figure}
\resizebox{88mm}{!}{\includegraphics{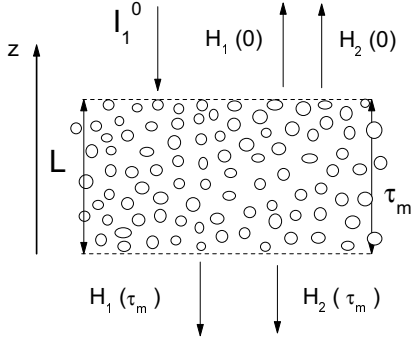}} \vspace{-20mm}
\caption[]{The geometry of the radiation transfer in a cloud slab.}
\label{rad_trsf}
\end{figure}

In this section we consider the radiation transfer   in the
cloud slab quantitatively. For simplicity we make the following
approximations:

1) The cloud slab is plane, homogeneous, infinitely extended in the x-y plane
and has a finite geometrical thickness $L$ in the z-direction (see Fig.
\ref{rad_trsf}).
This means that the radiation transfer is in z-direction.

2) We consider radiation transfer in two spectral bands only.
In band 1, that includes the soft X-ray and far UV bands, the slab
is illuminated from the outwards side.
Band 2 includes the optical (visual and soft UV) band.

3) We treat the reradiation of the incoming flux by one cloud as a
scattering. For single scattering the major part ($1-A$) of the flux in
band 1 is reradiated in the same band 1.
The smaller part ($A$) of the flux in band 1  is reradiated in band 2.
Incoming radiation in band 2 is fully reradiated in the same
band 2. We define the scattering coefficient on clouds as
\be
      \chi_c = \pi R_c^2 N_c~~{\rm cm}^{-1},
\ee
where $N_c$ is the cloud number density. The corresponding optical
depth increment is:
\be
     d\tau_c = - \chi_c~ dz.
\ee
In this case the radiation transfer equation in the two bands can be
written as:
\be
     \mu \frac{dI_{1,2}}{dz} = -(\chi_c +\chi_e) I_{1,2} + \eta^c_{1,2} +
     \eta^e_{1,2}.
\ee
We take electron scattering in the intercloud medium into account, too,
with the electron scattering coefficient $\chi_e$:
\be
      \chi_e = \sigma_T N_e~~{\rm cm}^{-1}.
\ee
Here $N_e$ is the electron number density of the intercloud medium,
$\eta^e_{1,2}$ and $\eta^c_{1,2}$ are the emissivities due to electron
scattering and cloud scattering in the first and second spectral bands.
One can  rewrite Eq. (3) as follows:
\be
    \mu \frac{dI_{1,2}}{d\tau_c} = (1+\alpha)I_{1,2} - S^c_{1,2} - S^e_{1,2},
\ee
where $\alpha=\chi_e/\chi_c$, and $S^c_{1,2}$, $S^e_{1,2}$ are the cloud
scattering and electron scattering source functions in both spectral bands.

4) We take isotropic electron scattering, but we can not consider the
cloud scattering as isotropic, since the clouds have a larger
dimension than the light wavelength. Therefore, incoming flux is reflected
mainly in the opposite direction, and the cloud scattering is described by
a redistribution function that is unknown. We treat the radiation
transfer in the two-stream approximation
(Mihalas \cite{Mihalas}, Suleimanov et al. \cite{SlMM})
for the directions $\mu_{1,2} = \pm 1/\sqrt{3}$, and
introduce the anisotropy function $\beta$ to take unisotropic scattering into
consideration. This function  describes the fraction of the radiation
reflected in opposite direction to the source function. The source functions
can be rewritten in this case as:
\be
        S^{c,\pm}_1 = (1-A) (\beta I_1^{\mp}+(1-\beta)I_1^{\pm})
\ee
\be
       S_2^{c,\pm} = A(\beta I_1^{\mp} +(1-\beta)I_1^{\pm}) +\beta I_2^{\mp}
       +(1-\beta) I_2^{\pm}
\ee
\be
      S_{1,2}^{e,\pm} = (I_{1,2}^{\pm} + I_{1,2}^{\mp})/2
\ee
The value of $\beta$ ranges from 0.5 to 1. The cloud scattering is isotropic
at $\beta$=0.5, and at $\beta$=1 the opposite directed radiation contributes
to the source function.

In the framework of the above mentioned approximations the
radiation field is described by 4 equations:
\be
    \pm \frac{1}{\sqrt{3}}\frac{dI_{1,2}^{\pm}}{d\tau} = (1+\alpha)
I_{1,2}^{\pm} - S_{1,2}^{c, \pm} - \alpha S^{e, \pm}_{1,2}.
\ee
Here and below we do not write the index c at $\tau$. Introducing the
mean intensity:
\be
     J_{1,2} = \frac{1}{2} (I_{1,2}^+ + I_{1,2}^-),
\ee
and the Eddington flux:
\be
     H_{1,2} = \frac{1}{2\sqrt{3}} (I_{1,2}^+ -I_{1,2}^-),
\ee
we rewrite Eq.(9) in the form:
\be
     \frac{dH_1}{d\tau} = A J_1
\ee
\be
     \frac{dJ_1}{d\tau} = 3(\alpha + A + 2\beta(1-A))H_1
\ee
\be
     \frac{dH_2}{d\tau} = - AJ_1
\ee
\be
     \frac{dJ_2}{d\tau} = 3((\alpha +2\beta)H_2 + A(2\beta-1)H_1)
\ee

Solution of equations (12)-(15) gives:
\be
     J_1 = C_1 \exp(-b\tau) + C_2 \exp(b\tau)
\ee
\be
     H_1 = \frac{d}{\sqrt{3}} (C_2 \exp(b\tau) - C_1 \exp(-b\tau))
\ee
\be
     H_2 = C_3 - H_1
\ee
\be
     J_2 = C_4 +3(\alpha + 2\beta)\tau C_3 - J_1
\ee
where
$$
     b=(3A(\alpha +A +2\beta(1-A)))^{1/2},
$$
and
$$
     d= \left( \frac{A}{\alpha + A +2\beta(1-A)} \right)^{1/2}
$$

The constants $C_1$, $C_2$, $C_3$, $C_4$ are obtained from the upper boundary
condition at $\tau=0$ and from the lower boundary condition at $\tau=\tau_m$.
We consider two cases for the lower boundary condition. In the first
case there is no medium and source of radiation under a slab. This
case corresponds to a radiation transfer through a cloud along
the accretion disc rim. In the second case there is a homogeneous
medium under the slab. This medium reprocesses the radiation from
the first to the second band with the same efficiency as a cloud.
The second case corresponds to radiation transfer through a slab
above an accretion disc at a fixed disc radius.

\subsection{The cloud slab seen along the accretion disc}

In this case there is no incoming radiation
at the lower boundary from below in both spectral bands: $I_1^+(\tau_m)=
I_2^+(\tau_m)=0$. At the upper boundary there is no incoming radiation
in the second spectral band, $I_2^-(0)=0$, but there is incoming
radiation $I_1^-(0)=I_1^0$ in the first spectral band. This means that
at the  upper slab boundary there is the external irradiating
flux $H_1^0 =-I_1^0/2\sqrt{3}$. These boundary conditions are expressed as:
\be
    J_1(0) - \sqrt{3}H_1(0) = I_1^0
\ee
\be
    J_2(0) - \sqrt{3}H_2(0) = 0
\ee
\be
    J_1(\tau_m) + \sqrt{3}H_1(\tau_m) = 0
\ee
\be
    J_2(\tau_m) + \sqrt{3}H_2(\tau_m) = 0
\ee

The constants obtained using these boundary conditions are:
\be
    C_1 = \frac{I_1^0}{1+d} \left( 1-\frac{1}{1-B}\right),
\ee
\be
   C_2 = \frac{I_1^0}{1-d} \cdot \frac{1}{1-B},
\ee
\be
     C_3 = - \frac{I_1^0}{2\sqrt{3}+3(\alpha+2\beta)\tau_m},
\ee
\be
      C_4= I_1^0 + \sqrt{3} C_3,
\ee
where
$$
      B=\exp(2b\tau_m) \left( \frac{1+d}{1-d} \right)^2
$$

We consider 4 functions for describing the radiation emitted from the slab.
The first one is the reprocessing efficiency from first to second band:
\be
      \eta = \frac{2\sqrt{3}H_2(0)}{I_1^0}.
\ee
The second one is the albedo of the cloud slab in the 1-st spectral band:
\be
      \zeta = 1+\frac{2\sqrt{3}H_1(0)}{I_1^0}.
\ee
The third and the fourth ones are the ratio of the flux coming out from the
lower boundary in the second and  in the first band, respectively, to the
ingoing flux:
\be
      \eta_1 = \frac{2\sqrt{3}H_2(\tau_m)}{I_1^0}.
\ee
\be
      \zeta_1 = -\frac{2\sqrt{3}H_1(\tau_m)}{I_1^0}.
\ee

It is possible to consider two limiting cases:

1) The cloud slab is absent ($\tau_m=0$). In this case there is no
reflected flux in the second band, and the flux in the first band is
equal to the incoming flux:
$$
    H_1(0) = - \frac{I_1^0}{2\sqrt{3}}.
$$

2) The semi-infinite cloud slab ($\tau_m = \infty$). In this case $C_2=
C_3=0$, and the reprocessing efficiency is:
\be
      \eta=\frac{2d}{1+d}.
\ee
This is the maximum possible value at given $A, \alpha$, and $\beta$.

\begin{figure}
\resizebox{88mm}{!}{\includegraphics{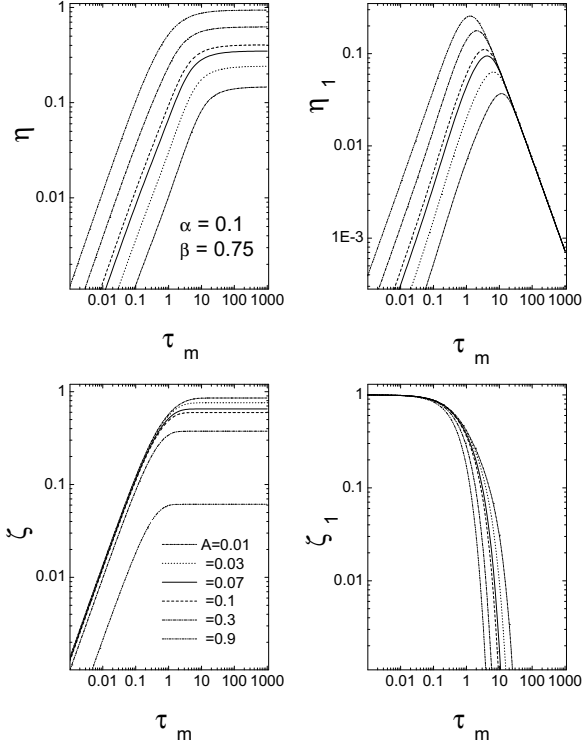}}
\caption[]{The functions $\eta$, $\zeta$, $\eta_1$, $\zeta_1$ vs. a cloud slab
optical depth $\tau_m$ at fixed $\alpha=0.1$ and $\beta=0.75$ and different
values of $A$.}
\label{a_dif}
\end{figure}

We now consider the dependence of $\eta$, $\eta_1$, $\zeta$ and $\zeta_1$ on
the slab optical depth $\tau_m$ and on the parameters $A$, $\alpha$
and $\beta$. The dependence on $\tau_m$ at different values of $A$ is shown
in Fig. \ref{a_dif} for values $\alpha=0.1$ and $\beta=0.75$. As expected
at small $\tau_m$ both reflected and transmitted   slab fluxes
in the second band are small, and in first band nearly all incoming flux is
transmitted through the slab ($\zeta_1 \approx 1$). At $\tau_m \approx 1$
photons on average are scattered once on clouds and, therefore,
$\eta \approx A$. The flux emitted in the second band from the
backside of the slab is also maximal at $\tau_m \approx 1-10$. At the largest
$\tau_m$  $\eta$ grows up to its maximum possible value (Eq. 32),
and $\zeta=1-\eta$.
This means that the rest part of the flux is reflected in the first band.
The fraction $\zeta_1$ of the transmitted radiation through the slab in the
first band $\zeta_1$ decreases exponentially with $\tau_m$ increasing, and
$\eta_1$ decreases when $\tau_m$ increases.
It is obvious that $H_1(\tau_m) \approx 0$ at large $\tau_m$ and
therefore $H_2(\tau_m) \approx C_3(\tau_m)$. This gives:
\be
     \eta_1 = (1+\frac{\sqrt{3}}{2}(\alpha + 2\beta)\tau_m)^{-1}.
\ee

\begin{figure}
\resizebox{88mm}{!}{\includegraphics{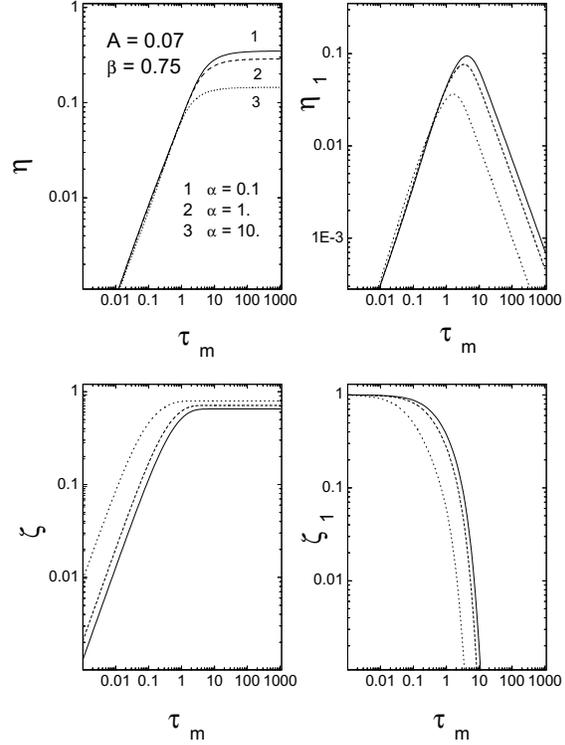}}
\caption[]{The functions $\eta$, $\zeta$, $\eta_1$, $\zeta_1$ vs. a cloud slab
optical depth $\tau_m$ at fixed $A=0.07$ and $\beta=0.75$ and different
values of $\alpha$.}
\label{alf_dif}
\end{figure}

Electron scattering decreases the reprocessing efficiency (see Fig.
\ref{alf_dif})
($\eta$ and $\eta_1$ decrease when $\alpha$ increases). The electron
scattering leads to reflection of the radiation without thermalization and
if $\chi_e$ is larger than $\chi_c$ most   of the irradiating flux
is reflected back by intercloud electrons without significant scattering
between clouds. This leads to decreasing  $\eta$ and $\eta_1$.

\begin{figure}
\resizebox{88mm}{!}{\includegraphics{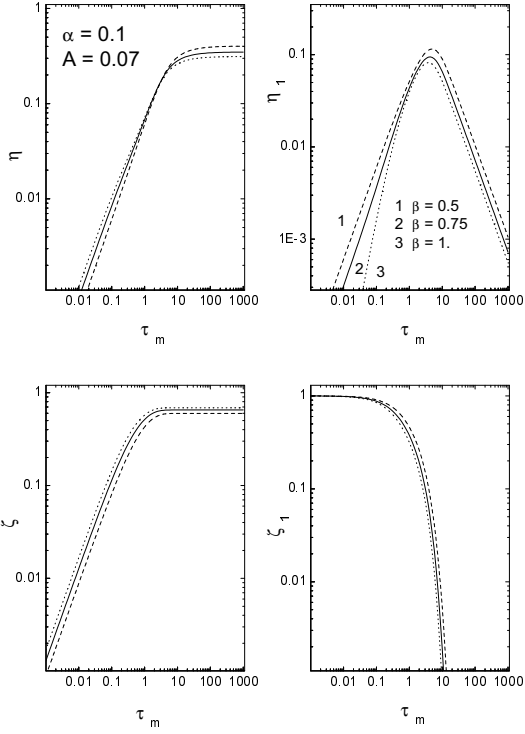}}
\caption[]{The functions $\eta$, $\zeta$, $\eta_1$, $\zeta_1$ vs. a cloud slab
optical depth $\tau_m$ at fixed $\alpha=0.1$ and $A=0.07$ and different
values of $\beta$.}
\label{bet_dif}
\end{figure}

The functions $\eta$, $\eta_1$, $\zeta$ and $\zeta_1$ depend only slightly
on the coefficient of anisotropy $\beta$ (Fig. \ref{bet_dif}). It is seen that
in the case of isotropic scattering on the clouds ($\beta=0.5$) the
reprocessing  efficiency is higher than in the case of unisotropic scattering
($\beta=1$). In unisotropic scattering the radiation is reflected
from clouds only backwards and penetrates less deeply into the
cloudy slab and therefore undergos a smaller number of scatterings
on clouds before escaping.

\subsection{The cloud slab seen above the accretion disc}

The reprocessing efficiency of the accretion disc is the same as for a cloud.
In this case the lower boundary conditions are expressed:
\be
           d_1^2 J_1(\tau_m) + \sqrt{3} H_1(\tau_m) = 0,
\ee
\be
           H_2 (\tau_m) + H_1(\tau_m) = 0,
\ee
where
\be
           d_1=(A/(2-A))^{1/2}.
\ee
Equation (35) leads to $C_3$=0, therefore $C_4=I_1^0$.

The upper boundary conditions are the same as in the case
considered before.

The constants $C_1$ and $C_2$ are:
\be
    C_1 = \frac{I_1^0}{1+d} \left(1+\frac{1-d}{1+d}\cdot \frac{1}{B_1}\right)
\ee
\be
   C_2 = \frac{I_1^0}{1+d} \cdot \frac{1}{B_1}
\ee
where
\be
 B_1 = \frac{1-d}{1+d} - \exp (2b\tau_m) \left(\frac{d_1^2+d}{d_1^2-d}\right).
\ee

\begin{figure}
\resizebox{88mm}{!}{\includegraphics{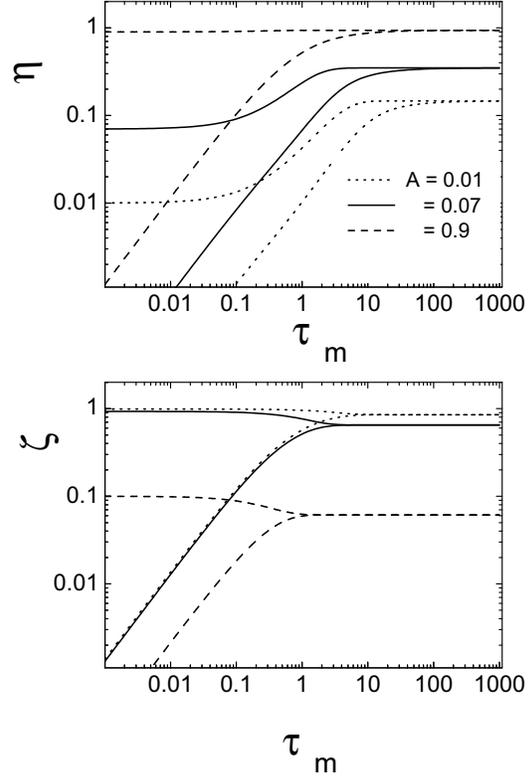}}
\caption[]{The functions $\eta$ and $\zeta$ vs. a cloud slab
optical depth $\tau_m$ at fixed $\alpha=0.1$ and $\beta=0.75$ and different
values of $A$ in the second case of the lower boundary conditions.}
\label{underd}
\end{figure}

It is obvious that $\eta \approx A$ if a cloud slab is absent ($\tau_m=0$)
and $\eta$ grows up to the value given by Eq. (32) if $\tau_m$ tends to
$\infty$. One notes that $\eta$ in the second case is greater
than in the first case at all $\tau_m$ (Fig. \ref{underd}).
Radiation  does not escape from the slab lower boundary, but
returns to the slab due to reflection from the accretion disc. As a result the
radiation emitted from the cloud slab has a larger number of scatterings
between clouds before escaping.

\section{Physical parameters of the clouds and the intercloud medium
for accretion discs in SSS}

There are several costraints on the clouds and the intercloud medium. In the
following we describe how the physical parameters are connected to the
properties of the matter in the spray, the disc, to the spray geometry,
the accretion rate and the radiation from the white dwarf.

(1) The mass flow rate in the spray has to be less than or equal
to that coming from the secondary star. Here the cross section of the
spray enters and the Kepler velocity. The modeling of Schandl et al.
(1997) for CAL 87 gives an example.

(2) For our suggestion of multiple scattering it is necessary that scattering
at clouds occurs more often than scattering in the intercloud medium. This is
a constraint on filling factor and cloud size for efficient reprocessing.

(3) The temperature difference between the inner irradiated side and the
cooler outer side of the spray allows to estimate the optical depth of the
cloud slab $\tau_m$.

(4) From the value of $\tau_m$ the scattering coefficient on the clouds
$\chi_c = \tau_m \ L$ ($L$ slab thickness) follows and that from the cloud
radius $R_c=\frac{3}{4} f/\chi_c$ with the filling factor $f$ of clouds in
the spray. Assuming a value for $f$ the density $\rho_c$ in the cloud follows
from the total amount of gas flowing through the spray.

(5) From a given cloud temperature and $\rho_c$ the optical depth in the
cloud $\tau_c$ follows. The value for $\tau_c$ should be large, $>1$,
becomes large if the opacity $\kappa_c$ is large enough. This is a constraint
for $\rho_c$. This can be reached more easily for a small filling factor.

(6) An important constraint is that the clouds remain confined in the
intercloud medium. The pressure in the clouds is related to the values
mentioned before in statement (4). For a given temperature
($ \approx 5\cdot 10^5$~K) in the intercloud medium pressure equilibrium
requires a certain density. The resultant electron scattering coefficient
$\chi_e$ should then be smaller than $\chi_c$ (statement (2)).

To study the physical parameters in the case of CAL 87 we took the values
for white dwarf mass, accretion rate, luminosity and disc size
from the light curve modeling of Schandl et al. (1997). We used a
slab thickness of $2 \cdot 10^{10}$~cm. For the temperature at the
non-irradiated side of the spray we took $ 20000$~K. With these values
we find a filling factor of about $\frac {1}{10}$ and the listed
constraints are fulfilled.

\section{Discussion}

In the first paper (Suleimanov et al. \cite{SlMM}) we have shown that the
reprocessing efficiency of the soft X-ray   to the optical band
due to irradiation is   low, $A$=0.05 - 0.1 for an accretion disc or
a star surface.

The optical depth of the cloud slab in the supersoft X-ray source
CAL 87 is large, documented by the small fraction of flux leaving the
back side of the spray compared to the irradiated flux at the illuminated
side. This means that the reprocessing efficiency in multiple scattering
of soft X-ray and far UV flux between clouds can be close to the maximum
possible value given by Eq. (32). In our example the electron
scattering in the intercloud medium is small in comparison with the
scattering on the clouds. Therefore, the reprocessing efficiency $\eta$
can reach 0.3--0.5.
This is sufficient to explain the large observed optical and UV
fluxes in the classical supersoft X-ray sources CAL 83 and CAL 87  (Popham \&
DiStefano \cite{PDS}).

It is obvious that the model considered here  is very simple, but it
reflects the main physical property of radiation transfer between clouds
above an accretion disc in SSS. The model suggested here can be further
developed by considering the radiation transfer in the cloud slab
using numerical methods that allow to leave the
two-stream approximation and to consider a more realistic redistribution
function for scattering on clouds.
Moreover, the height of the cloud slab can be comparable with the radial
thickness, therefore and one may consider the two dimensional
radiation transfer problem.
Cloud properties can depend on the location in the slab, particularly,
the cloud reprocessing efficiency
$A$
in the center of the slab can be
different from the efficiency of a cloud near the slab boundary. One may
consider this point, too.

\section{Conclusions}

We suggest that above the outer accretion disc in supersoft X-ray sources there
is a spray of geometrically small clouds. In this layer cool clouds
($T_c$ around 20000 K) are embedded in a hot intercloud medium
($T_e \approx 5\cdot 10^5$ K). This rim structure is irradiated by the soft
X-ray/far UV flux of a central hot white dwarf
($T_{eff} \approx 5\cdot10^5$ K). We show that in multiple
scattering between the clouds the reprocessing efficiency  of
soft X-ray to optical flux can increase significantly in comparison
with the reprocessing efficiency in simple scattering on a cloud.

In the framework of a simple analytical model we obtain a solution of
the radiation transfer problem in a plane slab of clouds irradiated by a
soft X-ray flux. We treat the reradiation of the flux impinging
on a cloud as a scattering. In single
scattering a fraction
$A$
($A \approx 0.05-0.1$) of soft X-ray or
far UV flux is reprocessed to the optical/soft UV flux and the rest part
($1-A$) is reflected in the far UV spectral band. The
radiation in optical/soft UV band is scattered in the
same band. The radiation reflected from the slab has been
scattered many times between clouds before escaping, and as a
result the fraction of the optical/soft UV radiation in the reflected
flux reaches 0.3-0.5. This explains the observed large optical and UV fluxes in some SSS by
reradiation of the white dwarf soft X-ray radiation. This is the main
conclusion of our work. The result establishes the spray as an
important element in the physics of SSS.

\begin{acknowledgements}
 VS has been supported by the Russian Basic Researches Fund
(grant 02-02-17174).
\end{acknowledgements}


\begin{thebibliography}{99}
\bibitem[1982]{MC} Mason K.O., Cordova F.A., 1982, ApJ, 262, 253
\bibitem[1978]{Mihalas} Mihalas D., 1978, Stellar atmospheres.
Freeman, San-Francisco
\bibitem[1978]{Milgrom} Milgrom M., 1978, AJ 208, 191
\bibitem[1996]{PDS} Popham R., Di Stefano R. 1996. In: Greiner J. (Ed.)
Workshop on Supersoft X-ray Sources, Garching, 1996, Lecture Notes in
Physics No 472. Springer Verlag, p.65
\bibitem[1997]{SMM} Schandl S., Meyer-Hofmeister E., Meyer F.
1997, A\&A, 318, 73
\bibitem[1999]{SlMM} Suleimanov V., Meyer F., Meyer-Hofmeister E.
1999, A\&A, 350, 63
\bibitem[1995]{WNP} White N.E., Nagase F., Parmar A.N., 1995, X-Ray Binaries,
eds. W.H.G. Lewin, J. van Paradijs, E.P.J. van den Heuvel, Cambridge
Astrophysics Series, p. 1
\end{thebibliography}
\end{document}